# Stochastic highway capacity: Unsuitable Kaplan-Meier estimator, revised maximum likelihood estimator, and impact of speed harmonisation

Igor Mikolasek

Transport Research Centre CDV, Lisenska 33a, 636 00 Brno, Czech Republic
igor.mikolasek@cdv.gov.cz

**Abstract:** The Kaplan-Meier estimate, also known as the product-limit method (PLM), is a widely used non-parametric maximum likelihood estimator (MLE) in survival analysis. In the context of highway engineering, it has been repeatedly applied to estimate stochastic traffic flow capacity. However, this paper demonstrates that PLM is fundamentally unsuitable for this purpose. The method implicitly assumes continuous exposure to failure risk over time – a premise invalid for traffic flow, where intensity does not increase linearly, and capacity is not even directly observable. Although parametric MLE approach offers a viable alternative, its earlier derivation for this use case suffers from flawed likelihood formulation, likely due to attempt to preserve consistency with PLM. This study derives a corrected likelihood formula for stochastic capacity MLE and validates it using two empirical datasets. The proposed method is then applied in a case study examining the effect of a variable speed limit (VSL) system used for traffic flow speed harmonisation at a 2-to-1 lane drop. Results show that the VSL improved capacity by approximately 10 % or reduced breakdown probability at the same flow intensity by up to 50 %. The findings underscore the methodological importance of correct model formulation and highlight the practical relevance of stochastic capacity estimation for evaluating traffic control strategies.

## 1 Introduction and literature review

Highway capacity is a fundamental variable in road traffic engineering representing the maximum number of vehicles that can traverse a road segment over a given period. A detailed discussion of capacity definitions is provided in Section 2. When demand exceeds capacity, congestions arise, leading to queue formation. Understanding the capacity characteristics of a highway or freeway is crucial for addressing numerous traffic engineering challenges, such as evaluating the adequacy of existing infrastructure, predicting congestions, or optimizing traffic modelling and control to prevent capacity drops due to the onset of congestion.

Recurring congestions and queues form at bottlenecks – points in the traffic infrastructure where capacity is reduced compared to surrounding sections. The bottleneck responsible for initiating congestion is called an active bottleneck. While multiple bottlenecks may exist along a road segment, only one active bottleneck causes a specific instance of congestion. Although freeway segments are rarely bottlenecks, so-called phantom congestions can form along extended sections, particularly during periods of high demand. Nevertheless, freeway congestions typically arise at specific points with reduced capacity, such as steep inclines, on- or off-ramps, lane drops, or work zones. Therefore, capacity measurements are usually focused on these bottlenecks with recurring congestions. In some cases, a moving bottleneck, such as a particularly slow vehicle, may occur. Notably, artificial moving bottlenecks can be employed as a traffic control measure to dissipate queues at stationary bottlenecks (Čičić & Johansson, 2022).

Traditionally, capacity has been regarded as a fixed value specific to a given bottleneck or road section. This perspective dates to Greenshields (1935) and the first traffic flow (TF) model which clearly defined maximum intensity (i.e. capacity) corresponding to certain TF speed and density. This fixed, deterministic capacity approach is very intuitive and user-friendly, making it appealing to practitioners and policymakers. Consequently, it remains widely adopted in practice. Throughout this paper, the term "(traffic flow) intensity" is used instead of "flow," as that can be ambiguous, referring either to the variable or the traffic process.

Later, the phenomenon of capacity drop was identified and confirmed by numerous studies (e.g. Banks, 1990; Cassidy & Bertini, 1999; Chung et al., 2007; Gazis & Foote, 1969; Hall & Agyemang-Duah, 1991; Srivastava & Geroliminis, 2013; Zhang & Levinson, 2004). This led to the distinction between pre-queue flow (PQF) and queue discharge flow (QDF). PQF is measured before a breakdown occurs and can reach higher values, whereas QDF is measured downstream from an active bottleneck after the breakdown and capacity drop have occurred. This makes breakdown prediction and modelling especially relevant as once congestion occurs the capacity drop makes it even more difficult to resolve.

Finally, the concept of stochastic capacity and the application of the Kaplan-Maier estimator (Kaplan & Meier, 1958), also known as product limit method (PLM), in the context of traffic flow began to appear in the 1980s (Hyde & Wright, 1986; Van Toorenburg, 1986), as noted by Minderhoud et al. (1997). Since then, the stochastic approach has gained increasing interest, with researchers adopting both PLM and other methods (Arnesen & Hjelkrem, 2018; Brilon et al., 2007; Chow et al., 2009; Elefteriadou et al., 1995; Lorenz & Elefteriadou, 2001; Shao, 2011; Weng & Yan, 2016).

According to Geistefeldt and Brilon (2009), the early papers that applied PLM to TF (Minderhoud et al., 1997; Van Toorenburg, 1986) made a fundamental mistake in its application. They incorrectly considered the breakdown flow to be any flow measured downstream from an active bottleneck, i.e. the queue discharge flow. While QDF can be considered as a measure of capacity by certain definitions, it is incompatible with the PLM because the TF is already congested when QDF is measured. To illustrate, it would be akin to recurrently including deceased patients in a lifetime analysis. Each breakdown should be associated with only one record of breakdown flow.

Some of the cited papers also use parametric maximum likelihood estimation (MLE) to estimate stochastic capacity. Crucially, both methods (PLM and the parametric MLE) were assumed to be practically "identical", except that one is non-parametric, since PLM can be derived as a non-parametric MLE. This was seemingly successfully proven by Brilon et al. (2005), who concluded that both yield corresponding results. However, the identity assumption does not hold due to a violation of PLM assumptions when applied to traffic flow. This paper shows that the corresponding results were found only due to an erroneous likelihood formula of the parametric MLE. See Sections 3.2 and 3.3 for more details.

This study originates from an applied research project aimed at developing a model to predict queue onset and evolution, capable of both long- and short-term predictions. An aggregation interval of 5-15 minutes has been commonly used in the existing literature, with TF intensities typically grouped into intervals. However, this aggregation reduces both the accuracy and resolution in terms of time and TF intensity. To address this limitation, a more disaggregated approach was adopted to better capture TF fluctuations and extremes that can lead to breakdowns, even when the average intensity is low. The PLM was initially selected for capacity estimation based on the literature. During development and validation, however, it became apparent that the estimated breakdown probability distributions consistently failed to reproduce the empirical cumulative frequencies of breakdowns ($CF_B$). Specifically, the breakdown probability was underestimated at low intensities and overestimated at high intensities. This discrepancy led to original research which uncovered the reasons for it and allowed to derive a corrected parametric MLE for stochastic capacity, presented in this paper.

Accurate estimation of stochastic capacity is essential for intelligent transportation systems (ITS) and modelling applications that do or could utilise it (e.g. Cicic et al., 2020; Elefteriadou et al., 2011), as well as for reliable evaluation and comparison of different traffic management strategies, such as speed harmonisation through variable speed limits (VSL), as shown in this paper. However, the apparent limitations of PLM and the original MLE mean that these methods may not be suitable for those purposes. Therefore, the findings in this paper are highly relevant for development of existing and future traffic control strategies. Additionally, this paper takes a thorough look at the definitions of capacity and describes a method for processing raw data suitable for similar stochastic capacity estimation methods, based on the classification framework proposed by Brilon et al. (2005). This method can be adapted by modifying the specific parameters to better suit different locations and application objectives.

The structure of the paper is as follows: Section 2 discusses the definition of highway capacity; Section 3 presents the data and data processing, the PLM and MLE methods, comparison and validation methodology, and some options for transforming the results; Section 4 provides the results of the method comparison, and of a case study on work zone capacity with and without speed harmonisation; Section 5 discusses the results and draws conclusions.

## 2 Highway capacity definition

The simplest and most general definition of highway capacity, as mentioned in the Introduction, is the maximum TF intensity achievable at the specific highway profile for which it is determined. While this definition is straightforward and easily understood at first glance, it raises several unanswered questions upon closer examination.

The Highway Capacity Manual (HCM) defines capacity as “the maximum sustainable hourly flow rate at which persons or vehicles reasonably can be expected to traverse a point or uniform segment of a lane or roadway during a given time period under prevailing roadway, traffic, environmental, and control conditions. Reasonable expectancy is the basis for defining capacity. A given system element’s capacity is a flow rate that can be achieved repeatedly under the same prevailing conditions, as opposed to it being the maximum flow rate that might ever be observed. Since the prevailing conditions (e.g. weather, mix of heavy vehicles) will vary within the day or from one day to the next, a system element’s capacity at any given time will also vary” (Transportation Research Board, 2016).

It is clear, not only from the citation, that the “highest achievable intensity” does not necessarily mean the highest value of TF intensity ever recorded. Such a value would not be practical for traffic engineering applications. Therefore, “maximal sustainable flow” or similar definitions are often used. Additionally, breakdowns occur at various TF intensities at the same locations and the QDF also varies over time, which implies that capacity is stochastic in nature. Many researchers agree on this (see Introduction) and the HCM recognises it, too. Additionally, the length of the aggregation interval plays a crucial role, as during PQF the intensity is defined by demand. Using a long interval can thus lead to lower average intensity and lower estimated capacity while in fact the road was not operating at capacity during the whole interval (i.e. the TF was not saturated).

Hence, there are at least four additional parameters associated with the capacity definition. Different capacity definitions can lead to very different results as they can be fundamentally different (e.g. pre-breakdown vs. post-breakdown capacity). The four parameters are:

- Aggregation interval (e.g. 1, 3, 5, 15, 60 minutes)
- Pre-breakdown capacity (e.g. max. PQF) vs. post-breakdown capacity (e.g. mean QDF)
- Stochastic capacity (e.g. QDF distribution) vs. single-valued capacity (e.g. mean QDF)
- Definition of "maximum" or "reasonable expectancy" in the case of single-valued capacity (e.g. mean QDF, 95$^{th}$ percentile of the PQF)

A clear definition of TF breakdown is also necessary, especially when dealing with pre-breakdown capacity. The definition of breakdown in a work zone with a 2-to-1 lane drop (sudden speed drop) can be different from the definition of a breakdown near an on-ramp merge on a 4-lane freeway (more subtle speed drop). Using unsuitable definitions to identify breakdowns may lead to incorrect conclusions.

There is no single "correct" all-encompassing definition of highway capacity or TF breakdown. Different definitions are suitable for different applications. When one is interested in predicting development of a queue length, a suitable definition of capacity could be mean QDF or its probability distribution. In either case, the aggregation interval must be specified, too. Alternatively, when one is trying to predict breakdowns and congestions, they need to focus on PQF and pre-breakdown capacity. The choice of aggregation interval is of much higher importance in this case. Very short interval can be too noisy as it is virtually impossible to exactly identify the direct cause and moment of a breakdown. A longer aggregation period smooths the extremes and is thus more robust and reliable. On the other hand, a too-long interval will inevitably include intervals with unsaturated flow, reducing the average TF intensity over the aggregated interval and shifting the distribution to the left. A shorter aggregation interval is generally more precise but prone to error, while a longer interval brings less resolution but is more robust. See Elefteriadou & Lertworawanich (2003) for further discussion about capacity and breakdown definitions.

In this paper, capacity is defined as a three-minute TF intensity directly leading to a breakdown (i.e. breakdown flow). The resulting capacity estimate is defined by a cumulative distribution function (CDF) of capacity or, conversely, by a TF survival function. Different aggregation periods were considered, but the three-minute interval seemed to be the best trade-off between the discussed pros and cons of the different interval durations. In any case, although the results would differ, the described methods are in principle independent of the aggregation interval and TF breakdown definition. Different intervals and/or breakdown definitions may be used if considered more suitable for the problem at hand.

# 3 Methods

## 3.1 Data measurement and processing

The data used in this study come from pilot testing of the ZIPMANAGER system, a mobile modular telematics system for speed harmonisation and congestion warning ahead of freeway work zones. It is an evolution of ViaZONE system (Ščerba et al., 2015) and consists of several profiles with non-intrusive TF detectors and variable LED traffic signs that automatically switch between pre-defined schemata based on current TF conditions. These signs function as variable speed limits and, when congestion occurs, as congestion warning.

The pilot test was conducted from September to November 2016 on the D5 freeway from Pilsen to Prague in Czechia. The system was installed between km 36.22 and km 31.30 ahead of a work zone with two-to-one lane drop. The default speed limits were 80 km/h inside and up to 100 m ahead of the work zone, 100 km/h from 100 to 500 m ahead, and 130 km/h elsewhere.

When active, the system could override these speed limits using three sets of LED traffic signs located approximately 0.9, 1.7, and 2.5 km ahead of the work zone, gradually reducing the limits to 120, 100, or 80 km/h. Three different speed harmonisation schemata were used, each with increasingly restrictive speed limits, aimed at increasing capacity and reducing risk of accidents. The signs were switched off during low traffic conditions.

The data used in this study come from a Wavetronics detector located approximately 100 m ahead of the lane drop. Measurement periods of 19 and 20 days (without and with the system active, respectively) were used for the case study. Given the average annual daily traffic of around 40,000 veh/day per both directions, queues were forming recurrently ahead of the work zone (52 and 39 identified breakdowns, respectively).

The choice of the data source has several advantages. The location allows almost immediate detection of queue at the lane drop. Only a few vehicles coming to a halt at the merging point while failing to properly zip merge can cause TF breakdown and congestion onset. The speed decreases rapidly from free flow to low speed or stop-and-go when congestion occurs at such a location. That sudden speed drop allows for relatively straightforward breakdown detection and reliable identification of the breakdown time and corresponding TF.

The detector itself is dependable in all weather conditions and, when set up optimally, can reach up to 99-% reliability in free flow conditions. Like most detectors, it starts failing in heavily congested conditions. However, that is not an issue as the data from congested periods are not utilised by the capacity estimation methods used in this study.

The data were processed into breakdown (non-censored), non-breakdown or free flow (censored), and discarded (congested, low-intensity, or inconclusive) overlapping intervals. The details of the processing, classification, and various thresholds are described in Appendix A. The use of overlapping intervals helps preserve information from the intermediate time steps, enabling more accurate identification of the moment of breakdown and the corresponding breakdown flow. However, this overlap must be kept in mind when interpreting the estimated capacity distribution. While the TF must be considered to be at risk of breakdown each minute, the breakdown probability is determined by the TF intensity over the past three minutes.

The final processed empirical datasets used in the case study include a total of 7,447 records (52 of which were breakdown flow, the rest is censored) without harmonisation, with TF intensities ranging from 46 to 112 PCE/3min, and a total of 9,565 records (39 breakdowns) with harmonisation, ranging from 46 to 114 PCE/3min.

### 3.2 Product limit method

The Kaplan-Meier estimator (PLM) is widely used tool in survival (or lifetime) analysis. Survival analysis focuses on estimating survival and failure probabilities, expected lifetimes, and related statistics. The survival function $S(t)$ is the complement of the cumulative distribution function $F(t)$ (eq. (1)). While the cumulative distribution function (CDF) represents the probability of failure before a given time $t$, the survival function describes the probability that the system "survives" longer than $t$.

$$F(t) = 1 - S(t) \tag{1}$$

The PLM can be used to estimate the survival function by incorporating both observed lifetimes and so-called censored data – cases where the subject did not "fail" during the observation period. Including censored data can significantly improve the accuracy of the estimate, especially when censored observations constitute a large portion of the dataset, as is the case in highway capacity analysis, where most TF data is censored. The estimated survival function $\hat{S}(t)$ is given by:

$$\hat{S}(t) = \prod_{j:\, t_j \leq t} \left(1 - \frac{b_j}{n_j}\right) \tag{2}$$

where $n_\mathrm{j}$ is number of records with lifetime $T \geq t_\mathrm{j}$ and $b_\mathrm{j}$ is the number of failures occurring at time $t_\mathrm{j}$. The Kaplan-Meier estimate is the product of partial survival probabilities over successive "age" intervals. By calculating the survival function through PLM, one can estimate the CDF via eq. (1).

**Table 1: The analogy between common survival (lifetime) analysis and its application to highway capacity analysis. Adapted from Brilon et al. (2005).**

| | Analysis of lifetime data | Capacity analysis |
|---|---|---|
| **Variable** | Time *t* | Traffic flow (TF) intensity *I* |
| **Failure event** | Death/failure at time *t* | Breakdown at TF intensity *I* |
| **Lifetime variable** | Lifetime *T* | Capacity *C* |
| **Censoring** | Lifetime *T* is longer than the duration of the experiment | Capacity *C* is greater than traffic demand |

According to Brilon et al. (2005), the method can be adapted for highway capacity analysis through analogy (Table 1). However, note that this method is only applicable to freeways or other dual-carriageway roads where each direction can be assessed independently and where vehicles have the right of way (i.e., there are no at-grade intersections except on- and off-ramps).

In this context, breakdowns occur when (PQF) capacity is exceeded. Therefore, the CDF which compliments the estimated survival function via (1) can thus be interpreted as the distribution of capacity and defines breakdown probability, effectively acting as a hazard function. This relationship was recognized by all the cited authors, who correctly used the CDF to define breakdown probability. Estimating the survival function via PLM and deriving the CDF was then a natural method.

However, this analogy overlooks a key mismatch – traffic flow does not exhibit any "aging". Simply put, TF intensity does not increase linearly, and capacity cannot be observed, causing the analogy to fail. Unlike subjects in survival analysis whose risk exposure accumulates over time, traffic flow intensity does not progress linearly, but fluctuates quasi-randomly around a time-varying mean. As a result, each TF observation acts as a distinct, independent "entity" that emerges at a specific intensity level, without a history of prior exposure to lower (or higher) intensities. This stands in contrast with traditional survival analysis, where each subject is continuously exposed to risk of failure throughout its lifetime, and the hazard is a function of time or age. Meanwhile, in the case of traffic flow, the hazard is a function of the TF intensity, which is randomly fluctuating, rather than progressing steadily. This is also why the capacity cannot be observed, unlike a lifetime – we can only infer it was lower than the current TF intensity. Note that this is unrelated to (right) censoring, which can and does happen in both cases. The difference is that with the right censoring we know that something has survived certain age or TF intensity, while the issue with capacity is that we cannot infer any lower bound. One could say that the "uncensored" data are actually left censored instead.

To some extent, this situation resembles estimating a material's strength distribution: failure occurs when the applied stress exceeds the material's (unobserved) strength, but only the applied load can be measured directly. Moreover, if the loading does not cause failure, the observation is treated as censored. In practice, the load is increased continuously, allowing the induced stress at failure to be estimated, and thus the strength of the failed sample determined. However, if the load were applied instantaneously in an impulse-like test, where each specimen was subjected to a single, known load with a pass/fail outcome, the situation would be analogous to the TF, although the practicality of such a test is dubious. Applying the PLM in this context would also be inappropriate.

This mismatch manifests followingly in the PLM calculation. According to eq. (2), the survival function is estimated through a product of partial survival probabilities, where the hazard at each age interval is estimated as $b_j/n_j$. It is important to recognize that this "traditional" hazard is fundamentally different from the "hazard function" that emerges from interpreting the CDF of capacity as breakdown probability. In the formula, $n_j$ represents the number of subjects still at risk at time $t_\mathrm{j}$, based on the assumption that subjects persist through time – if subject is alive at $t_\mathrm{j}$ when it is observed, they must have been alive at all prior times $t_{\mathrm{j-m}}$.

**Table 2: Excerpt from the PLM calculation table based on real data. The "Exposed to risk:" column compares the actual exposure to failure risk, defined as the number of TF records observed at a given intensity $I_\mathrm{j}$, with the exposure assumed by the PLM, defined as the number of TF records with intensity ≥ $I_\mathrm{j}$. Note that no breakdowns were observed at, e.g., TF intensities 57-59 PCE/3min, so $\widehat{S}(I_\mathrm{j})$ cannot be calculated via PLM for those, resulting in a grouped interval $I_\mathrm{j} = 56 - 59$. Values of $r_{\mathrm{I_j}}$ in brackets represent the total exposure over the entire interval between intensities with recorded breakdowns.**

| TF int. $(I_\mathrm{j})$ | No. of events $(b_\mathrm{j})$ | Exposed to risk: Reality $(r_{\mathrm{I_j}})$ | PLM $(n_\mathrm{j})$ | Part. failure probability | Part. survival probability | $\widehat{S}(I_\mathrm{j})$ |
|---|---|---|---|---|---|---|
| 56 | 2 | 298 (1159) | 6445 | 0.00031 | 0.99969 | 0.99945 |
| 60 | 1 | 318 (318) | 5408 | 0.00018 | 0.99982 | 0.99927 |
| 61 | 1 | 271 (271) | 5123 | 0.00020 | 0.99980 | 0.99907 |
| … | … | … | … | … | … | … |
| 105 | 1 | 5 (7) | 18 | 0.05556 | 0.94444 | 0.84784 |
| 108 | 1 | 2 (5) | 11 | 0.09091 | 0.90909 | 0.77077 |
| 112 | 1 | 2 (6) | 6 | 0.16667 | 0.83333 | 0.64231 |

This continuous exposure is essential to time-based hazard modelling. However, it does not hold true for the traffic flow due to the absence of "aging", as discussed previously. Applying the same logic to it then vastly overestimates the exposure to risk at lower TF intensities (see Table 2), leading to underestimation of the hazard (i.e. partial failure probability), overestimation of the survival function, and ultimately underestimation of the breakdown probability.

One might instinctively consider defining $n_j$ as the number of TF records observed at intensity $I_j$, but this introduces the opposite problem as it does not reflect the correlation between similar intensities. Each breakdown then has too much weight at its specific intensity level, leading to overestimation of the partial failure probability and these errors then propagate and compound due to the multiplicative nature of the PLM calculation.

On the other hand, PLM overestimates the breakdown probabilities at higher intensities due to the mismatch between what PLM calculates as a hazard (i.e. the partial failure probability) and what actually defines the breakdown probability, the capacity CDF. Using the estimated CDF results in too many breakdowns in higher intensities as the hazard is (indirectly, via eq. (2)) multiplied over all the levels. Conversely, using the actual estimated hazard function results in too few breakdowns as it is underestimated, particularly at low intensities. There is no way to use the outcomes of PLM to obtain accurate breakdown probabilities at specific TF intensity levels.

However, when the CDF derived from PLM is used, the total expected number of breakdowns across all intensities remains approximately correct (within the margin of sampling error). This makes the issue particularly difficult to detect. The shape of the CDF (and the corresponding cumulative frequency curve) also appears plausible. It is "just" steeper than it should be, making the distortion not immediately apparent through visual inspection.

### 3.3 Maximum likelihood estimator

An alternative approach to estimating breakdown probability or the capacity distribution is the parametric maximum likelihood estimator (MLE) (Myung, 2003). This method identifies the optimal parameter values of a chosen probability distribution by maximizing the likelihood function, defined as $L(\theta|\boldsymbol{y}) = f(\boldsymbol{y}|\theta)$, which represents the likelihood of the parameter vector $\theta$ given the observed data $\boldsymbol{y}$ (eq. (3)). The likelihood function is therefore specific to the particular problem and distribution under consideration.

$$\hat{\theta} = arg \max_{\theta \in \Theta} L(\theta, \boldsymbol{y}) \tag{3}$$

The MLE is a widely used method in statistics, including in the context of material strength distribution, especially for estimating fatigue resistance and stress-life (S–N or Wöhler) curves (Ambrožič & Gorjan, 2011; Pollak & Palazotto, 2009). It has also been proposed as a parametric alternative to PLM and directly compared to it by Brilon et al. (2005), which conclude that both methods yield comparable results, with differences attributable to the distinction between empirical and parametric curves. The paper also provides a formula specifically derived for the use case of capacity estimation:

$$L = \prod_{i=1}^{n} f_c\,(q_i)^{\delta_i} \cdot [1 - F_c(q_i)]^{1-\delta_i} \tag{4}$$

where $q_i$ denotes the TF intensity (flow) in the $i$-th observation interval, equivalent to $I_i$ in this paper; $f_c$ and $F_c$ represent the probability density function and cumulative distribution function of capacity, respectively; and $\delta_i$ is an indicator variable marking failure in the $i$-th observation (1 for failure, 0 for survival). This formula was applied in earlier stages of this study, as documented in a previous version of this paper (Mikolasek, 2020). However, since it yields results virtually identical to those of PLM, it should now be evident that its application also leads to incorrect results. Derivation of the likelihood expression from first principles reveals a flaw. The statistical function in both terms of the equation should, in fact, be $F_c(q_i)$.

The likelihood function is defined as the product of the likelihoods $L_i$ of all observations:

$$L = \prod_{i=1}^{n} L_i \tag{5}$$

In the context of traffic flow capacity, $L_i$ is given as the probability of a breakdown $P_B(\mathrm{I}_i)$ at the observed TF intensity $\mathrm{I}_i$ if the outcome is breakdown ($\delta_i$=1), or as the probability of "survival" $P_S(I_i)$ if the outcome is survival ($\delta_i$=0):

$$L_i = P_B(I_i)^{\delta_i} \cdot \left(P_S(I_i)\right)^{1-\delta_i} = P_B(I_i)^{\delta_i} \cdot \left(1 - P_B(I_i)\right)^{1-\delta_i} \tag{6}$$

Crucially, as discussed in Section 3.2, the breakdown probability function of TF is defined as the probability that the capacity $C$ is lower than the observed flow, which is defined by the CDF of capacity, therefore $P_B(I_\text{i}) = P(C < I_\text{i}) = F_c(I_i)$. From this, we can rewrite eq. (6) as:

$$L_i(\theta) = F_c(I_i)^{\delta_i} \cdot \left(1 - F_c(I_i)\right)^{1-\delta_i} \tag{7}$$

Substituting eq. (7) into eq. (5) yields the correct form of the likelihood formula:

$$L(\theta) = \prod_{i=1}^{n} \left[F_c(I_i)^{\delta_i} \cdot \left(1 - F_c(I_i)\right)^{1-\delta_i}\right] \tag{8}$$

Comparing this to eq. (4) as given by Brilon et al. (2005), we can see that the probability distribution function used in the original formulation to represent the likelihood of a breakdown has been replaced by the cumulative distribution function. As with PLM, for practical applications, the range of $i$ can be limited to the set of relevant intensities, i.e. $i \in \langle I_{min}, I_{max} \rangle$.

This error by Brilon et al. (2005) may stem from the authors' assumption that PLM was appropriate for the traffic flow context. As a result, they may have either (a) overlooked the mistake because the outcomes appeared "correct" or (b) derived the formula correctly, and, upon obtaining results that deviated from PLM, they searched for and ultimately introduced an error to reconcile the discrepancy. From there, the flawed formulation propagated into the subsequent literature, along with the PLM (Brilon et al., 2007; Geistefeldt & Brilon, 2009; Kianfar & Abdoli, 2021; Knoop & Hoogendoorn, 2022; Waleczek et al., 2016).

Since eq. (8) becomes numerically unstable for larger samples due to the likelihood asymptotically approaching zero, the log-likelihood is commonly used in practice:

$$\ell(\theta) = ln\, L(\theta) = \sum_{i=1}^{n} ln\, L_i \tag{9}$$

Based on the assumption that TF capacity follows Weibull distribution $W{\sim}(\lambda, \gamma)$, the best estimate of the capacity distribution can then be found by maximizing the log-likelihood by varying the parameters $\lambda, \gamma$ via eq. (10):

$$\hat{\theta} = [\lambda, \gamma]^T = \underset{\lambda,\gamma}{arg\,max}\, \ell = \sum_{i=1}^{n} \left[\delta_i \cdot ln\left(F_c(I_i)\right) + (1 - \delta_i) \cdot ln\left(1 - F_c(I_i)\right)\right] \tag{10}$$

where, by definition of the Weibull distribution:

$$F_c(I_i) = 1 - e^{-(I_i/\lambda)^\gamma} \tag{11}$$

The parameters $\lambda, \gamma$ define the scale and shape of the distribution, respectively.

### 3.4 Method comparison and validation

This section describes a validation method originally used to uncover the malfunction of PLM in the context of road capacity. It is used to compare the three methods discussed in this paper. The approach relies on the premise that a valid capacity model should, when applied to empirical TF data, correctly predict the expected number of breakdowns at each TF intensity level via eq. (12). To assess this, the cumulative frequency of breakdowns ($CF_B$, eq. (13)) is used to compare the predicted breakdowns against the empirical curve.

$$b_j = r_{I_j} \cdot F_C(I_j) \tag{12}$$

$$CF_B(I_i) = \sum_{j=I_{min}}^{I_i} b_j \tag{13}$$

Note that indexes $i$ and $j$ both indicate levels of TF intensity but are used to distinguish between different levels within the same equation. $CF_B(I_i)$ denotes the cumulative frequency of breakdowns at TF intensity $I_i$; $b_j$ is the number of breakdowns at intensity $I_j$, analogous to its use in PLM, either empirically observed or predicted via eq. (12); $r_{I_j}$ is the number of TF records observed at intensity $I_j$. In PLM terms, $n_j$ corresponds to the sum of $r_{I_j}$ from $I_j$ up to the highest recorded intensity $I_{max}$. Both censored and uncensored observations are included in $r_{I_j}$, while only the uncensored data (i.e. actual breakdowns) are included in $b_j$ for calculating the empirical $CF_B$ curve.

Different error metrics can be used to compare the estimated $CF_B$ curves $\widehat{CF}_B$ to the empirical benchmark. Commonly used methods include the sum of squared errors (SSE):

$$SSE = \sum_{i=I_{min}}^{I_{max}} \left(CF_B(I_i) - \widehat{CF}_B(I_i)\right)^2 \tag{14}$$

and the average relative error (ARE):

$$ARE = \frac{1}{n} \sum_{i=I_{min}}^{I_{max}} RE(I_i) \ , \qquad n = |\{I_{min}, I_{min} + 1, \dots, I_{max}\}| \tag{15}$$

where the relative error $RE$ is defined as:

$$RE(I_i) = \left| \frac{\left(CF_B(I_i) - \widehat{CF}_B(I_i)\right)}{CF_B(I_i)} \right| \tag{16}$$

However, for $CF_B$ curves, empirical breakdowns are typically sparse at low TF intensities, whereas the theoretical predictions gradually increase from zero. Due to the cumulative nature of $CF_B$, the relative error tends to remain close to 100 % at low intensities and gradually decreases towards higher intensities. To address this, a weighted version of the relative error can be used, where the weights are given by the expected number of breakdowns $\bar{b}_i$ at each TF intensity level, as defined in eq. (17). This average weighted relative error (AWRE) provides a more meaningful metric, reflecting the actual impact of estimation errors across the full range of TF intensities.

$$AWRE = \frac{1}{\sum_{i=I_{min}}^{I_{max}} \bar{b}_i} \cdot \sum_{i=I_{min}}^{I_{max}} \bar{b}_i \cdot RE(I_i) \tag{17}$$

Similarly, while the simple SSE is practical for optimisation, it can be replaced by the root mean square error (RMSE), defined as $RMSE = \sqrt{SSE/n}$. This yields a more interpretable absolute error metric and avoids the issue of diminishing relative errors for cumulative curves. Nevertheless, especially the weighted relative errors still have their place in model evaluation.

### 3.5 Transformation to longer intervals and expected time to breakdown

As Brilon et al. (2005) pointed out, the probability that TF remains in free flow for 60 minutes (assuming constant TF intensity) when the traffic flow is aggregated in non-overlapping 5-minute intervals, is $P_{S,60} = \left[1 - F_{C,5}(I_5)\right]^{\frac{60}{5}}$ or, in general:

$$P_{S,T} = \left[1 - F_{C,T_0}\left(I_{T_0}\right)\right]^{\frac{T}{T_0}} \tag{18}$$

This follows from eq. (1) and the principle that the probability of a random event within $n$ independent trials, each with probability $P$, is $P^n$, where $n = T/T_0$. Conversely, from eq. (1) again, the probability of breakdown occurring within the next $T$ minutes is given by:

$$P_{B,T}(I_i) = 1 - \left[1 - F_{C,T_0}\left(I_{T_0,i}\right)\right]^{\frac{T}{T_0}} = F_{C,T_0}\left(I_{T_0,i}\right)^{\frac{T}{T_0}} \tag{19}$$

For Weibull distribution, substituting eq. (11) into the first version of eq. (19) results in:

$$P_{B,T}(I_i) = 1 - \left\{\exp\left[-\left(\frac{I_{T_0,i}}{\lambda}\right)^{\gamma}\right]\right\}^{\frac{T}{T_0}} = 1 - \exp\left[-\frac{T}{T_0}\left(\frac{I_{T_0,i}}{\lambda}\right)^{\gamma}\right] \tag{20}$$

This study uses overlapping aggregation intervals, as described in Section 3.1. Importantly, the specific duration of the aggregation interval is not inherently critical – as long as it is consistent with the interval used to estimate the Weibull parameters. That is, if the parameters were estimated using 3-minute aggregated traffic flow data, then breakdown probability calculations must also be based on 3-minute intensities. This is because the scale parameter and the input intensity must remain proportional; any mismatch would invalidate the model.

What is relevant, however, is the breakdown evaluation frequency. Let $T_f$ denote the failure testing interval or period and $T_a$ the aggregation interval ($T_f = 1$ and $T_a = 3$ in this study). Still assuming constant traffic intensity over the evaluation period $T$, the breakdown probability can be calculated as:

$$P_{B,T}(I_i) = 1 - \exp\left[-\frac{T}{T_f}\left(\frac{I_{T_a,i}}{\lambda_{T_a}}\right)^{\gamma_{T_a}}\right] = F_{C,T_a}\left(I_{T_a,i}\right)^{\frac{T}{T_f}} \tag{21}$$

This is practically equivalent to eq. (19) and (20), with the distinction that the interval indexes $f$ and $a$ are explicitly stated. This helps clarify the difference between aggregation and failure testing intervals, which becomes relevant when these intervals differ.

Naturally, through eq. (1), the survival probability is given as:

$$P_{S,T}(I_i) = \left[1 - F_{C,T_a}\left(I_{T_a,i}\right)\right]^{\frac{T}{T_f}} \tag{22}$$

Since the breakdowns of uncongested TF can be considered as events in a Poisson point process (assuming constant intensity), the capacity distribution can be transformed into an exponential distribution to model the (mean) time to breakdown. The breakdown rate at given TF intensity $I_i$ is given by the capacity CDF $F_C(I_i)$, resulting in the time to breakdown $t_B(I_i) \sim \exp\left[F_C(I_i)\right]$. By the definition of exponential distribution, the mean time to breakdown is $E[t_B(I_i)] = 1/F_C(I_i)$. Since in this case the time is measured in number of failure tests, this value must be scaled by the length of the failure testing interval $T_f$ to convert it into real time (e.g., minutes). This results in:

$$E[t_B(I_i)] = T_f / F_C(I_i) \tag{23}$$

However, because the distribution of time to breakdown is positively skewed, the mean can overestimate the typical time before breakdown, making median potentially more useful.

$$Median[t_B(I_i)] = T_f \cdot \ln 2 / F_C(I_i) \tag{24}$$

Given the memoryless properties of uncongested TF and exponential distribution (which no longer holds once a breakdown occurs, as subsequent breakdowns are not possible until the current one has resolved), the time to breakdown provides an alternative mean of stochastic breakdown modelling. A random time to breakdown can be generated from the exponential distribution. If the time is less than the (modelled) time to the next expected change in TF intensity, a breakdown is considered to occur at that time. Otherwise, a new time to breakdown is generated at the start of the subsequent TF intensity interval.

# 4 Results and their interpretation

## 4.1 Comparison of the estimation methods

To compare the estimation methods, data from the period without traffic control (speed harmonisation via VLS) are shown, as it contains more breakdown events. The controlled period was also analysed, and the results were found to be qualitatively identical.

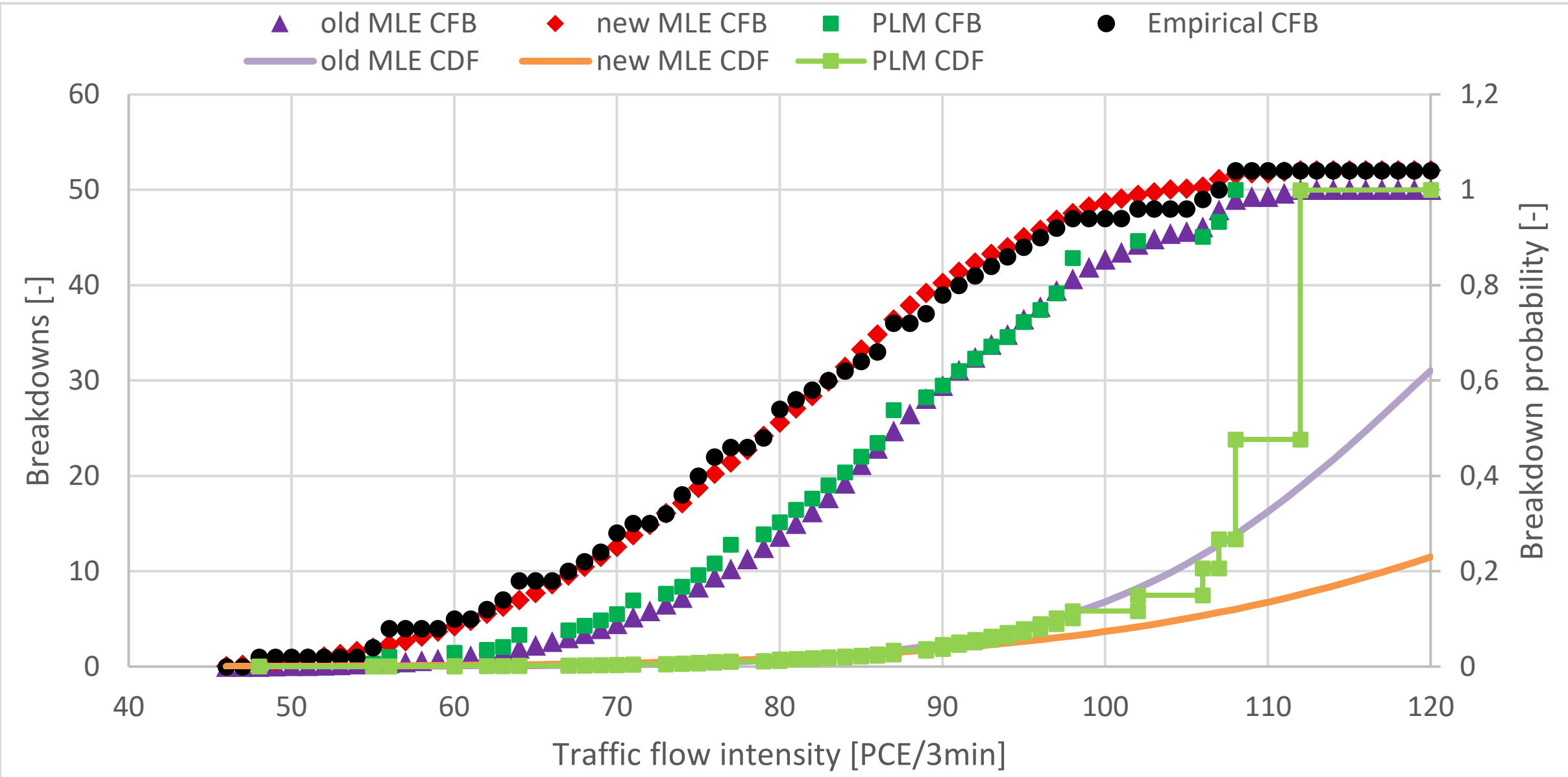

**Figure 1: Comparison of cumulative frequency of breakdown ($CF_B$) curves predicted by three capacity estimation methods (PLM, original MLE, new MLE) against empirical breakdown data. The underlying breakdown probability curves are also shown.**

Figure 1 presents a comparison of the $CF_B$ curves generated using the three estimation methods. These curves, shown in the central part of the chart and corresponding to the left axis, are based on the same traffic flow data and are compared against the empirical $CF_B$ curve, which serves as a benchmark. As expected, the curves derived from the PLM and the original MLE methods are similar; however, both deviate noticeably from the empirical curve. In contrast, the curve based on the capacity distribution estimated using the newly derived MLE method closely aligns with the empirical data.

The associated breakdown probability curves are shown in the bottom-right part of the chart, plotted against the right axis. At higher traffic intensities, both the PLM and original ("old") MLE methods clearly overestimate the breakdown probability compared to the new MLE.

Although the curves appear nearly identical at lower intensities, the underlying numerical values reveal the opposite – both old methods predict unrealistically low probabilities in this range, leading to the underestimation of low-intensity breakdowns in the resulting $CF_B$ curves. The estimated capacity distributions were $C \sim W(120.37;\ 10.38)$ with the old MLE and $C \sim W(146.42;\ 6.75)$ with the new, corrected one.

**Table 3: Error metrics of $CF_B$ curve predictions for different capacity estimation methods.**

| Capacity CDF estimation method | SSE | RMSE | ARE | AWRE |
|---|---|---|---|---|
| **Kaplan-Meier estimate (PLM)** | 2572 | 8.02 | 39.89 % | 30.46 % |
| **"Old" maximum likelihood (MLE)** | 3690 | 7.01 | 39.69 % | 32.56 % |
| **"New" maximum likelihood (MLE)** | 80 | 1.03 | 8.01 % | 5.87 % |

Error metrics for the breakdown predictions ($CF_B$ curves) were computed for all three methods, using the empirical data as ground truth. The results are summarised in Table 3 and align with the graphical comparison shown in Figure 1. The errors could not be calculated for the breakdown probability curve (i.e. the capacity CDFs) as the true underlying capacity distribution is unknown. For PLM, the errors were calculated only in points with known value of survival function, i.e. for intensities where a breakdown was recorded. The SSE is therefore divided by smaller $n$ when calculating RMSE, resulting in higher RMSE compared to the "old" MLE despite the lower SSE.

Both the visual and numerical results clearly demonstrate the limitations of the PLM when applied to traffic flow, and of the original MLE formula from Brilon et al. (2005). Although the new MLE still exhibits an average relative error of 8 %, or 6 % when weighted by relevance, these errors can be attributed to the random fluctuations in the empirical $CF_B$ curve due to the relatively small sample size. In contrast, the errors from the other two methods are clearly systematic. While the numerical error values quantify the differences, they are largely circumstantial due to the aforementioned; in this case, the graphical comparison of $CF_B$ curves might offer the most illustrative evidence.

## 4.2 Case study on real work zone data

Figure 2 presents the empirical $CF_B$ curves obtained from measurements during periods with and without the speed harmonization system (VSL) active, alongside the corresponding predicted curves. The underlying estimated capacity CDFs are also shown, illustrating the reduction in breakdown probability under VSL operation.

A notable "bump" of the empirical $CF_B$ curve with VSL is visible around 65 PCE/3min. There are a few possible causes of this, including pure randomness. However, it could also be an indirect consequence of the increased capacity enabled by the speed harmonisation. These intensities are commonly observed throughout the day and with the use of VSL can be sustained for longer periods. As a result, exposure increases and results in a disproportionally higher number of breakdowns at these intensities, illustrating diminishing returns of increasing capacity. This resembles the well-known traffic induction phenomenon observed when capacity is increased via adding more lanes (e.g. Hymel, 2019; Litman, 2017), which can eventually lead to more traffic and congestions, although, in this case, things happen on a much shorter time scale and the demand is already mostly determined, rather than induced. Nevertheless, the use of (mobile) VSL offers clear benefits by enabling more efficient utilization of (temporarily constrained) existing infrastructure and by helping to delay or prevent the onset of congestion at relatively low cost.

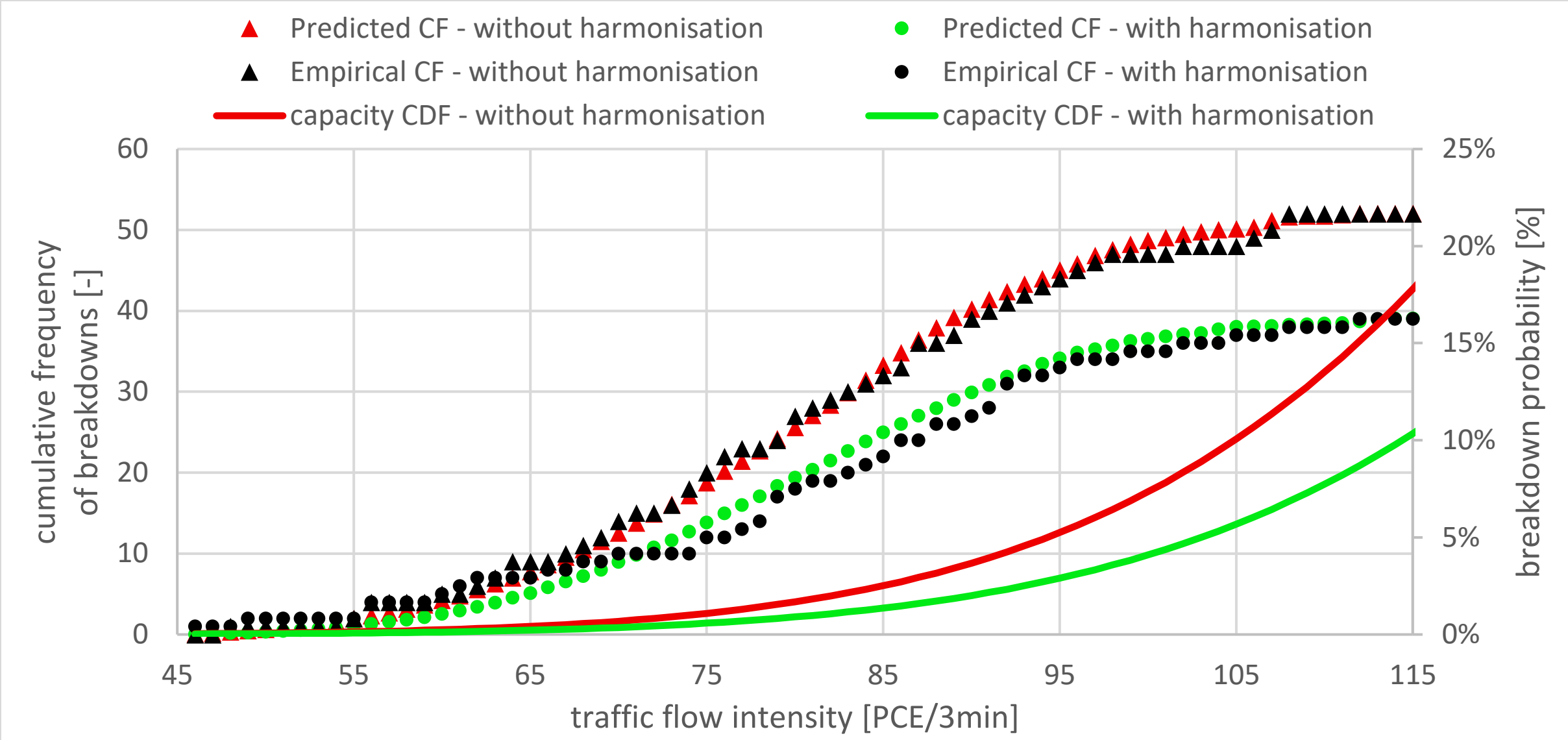

**Figure 2: Comparison of the empirical $CF_B$ curves (black, left axis) with and without harmonization, the estimated CDF curves (lines, right axis) and the $CF_B$ curves predicted from them (coloured, left axes).**

The increased probability of the congestion forming inside the work zone and spilling over due to the increased capacity of the merging zone is also a possible (co-)cause of the bump. In the case of a spillover, the recorded capacity is not the actual cause of the breakdown and is mostly coincidental. Since this range of intensities is the most common, it would make sense that proportionally more spilled breakdowns are recorded during those. Ideally, the spillback would be identified and treated separately, but that would require more detection profiles directly in the work zone, and, eventually, additional prediction model to incorporate them back, further increasing the overall model complexity.

The broad observed range of intensities at which breakdowns occur underscores the limitations of relying on a single-valued capacity for applications beyond basic layout design, although designing to provide sufficient capacity is also problematic due to the induced traffic leading to infinite loop of adding more lanes.

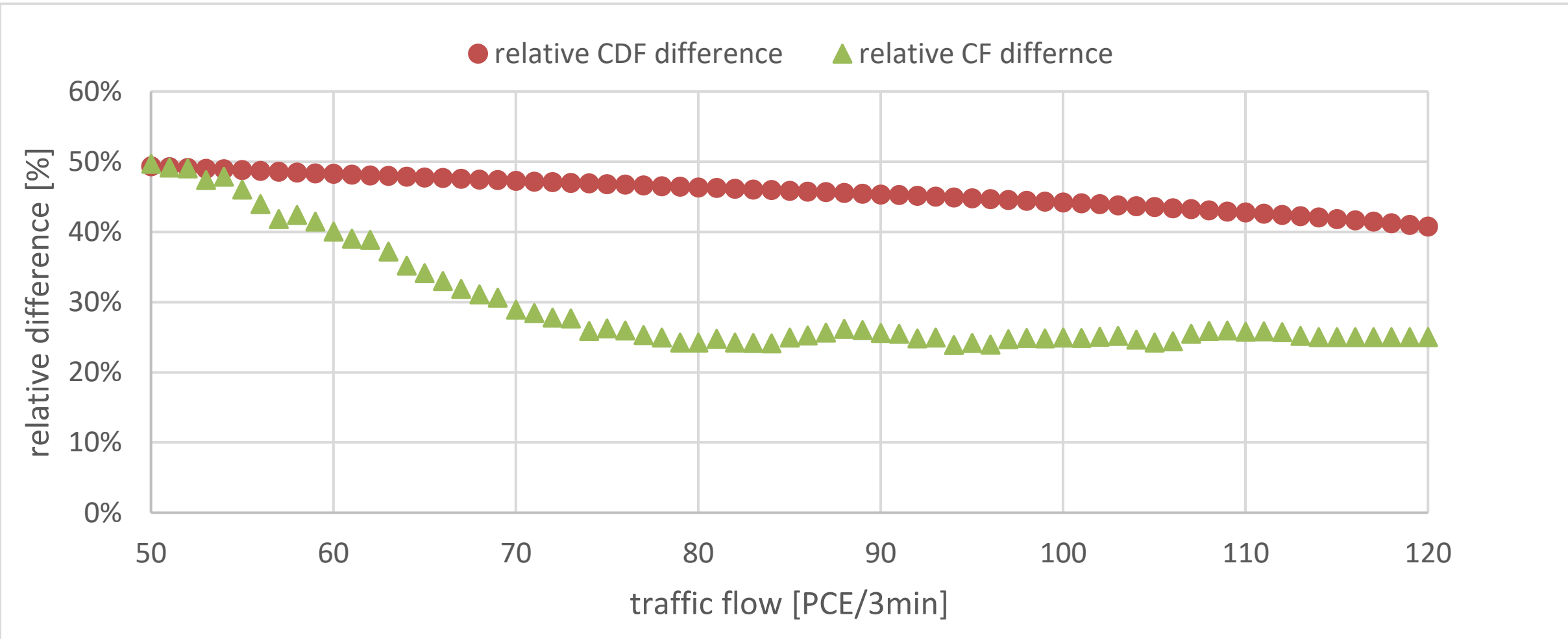

**Figure 3: Comparison of relative difference in breakdown probability and cumulative frequency of breakdowns with and without traffic speed harmonisation.**

The impact of harmonisation, as well as of the diminishing returns, is further illustrated in Figure 3. The breakdown probability decreases approximately linearly by about 40-50 % across the relevant intensity range, with a larger reduction at the lower end. However, this translates into much smaller relative decrease in the number of observed breakdowns, which gradually drops from the 50 % at low intensities around 50 PCE/3min to “only” about 25 % at intensities around and above 75-80 PCE/3min.

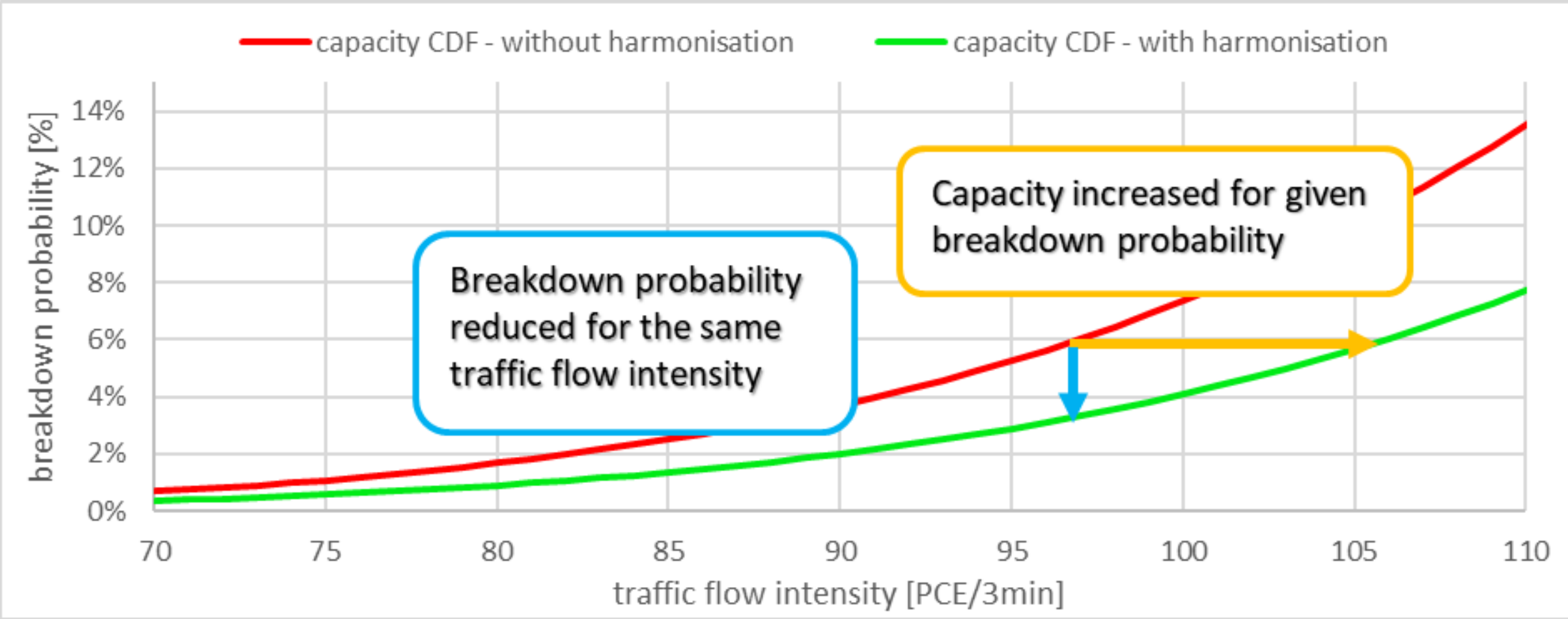

**Figure 4: Illustration of dual interpretation of increased capacity.**

The change in capacity can also be expressed as an increase in intensity corresponding to specific breakdown probabilities, represented by the left-to-right shift of the capacity CDF (Figure 4). This shift can be interpreted similarly to changes in the traditional deterministic capacity, which may be viewed as a stepwise CDF with a sudden jump from 0 to 1 at the capacity threshold. The shift was calculated at six different breakdown probability levels ranging from 0.1 % to 10 %, corresponding to the whole relevant range of intensities. On average, the difference amounted to 7.5 PCE/3min, equivalent to 150 PCE/h. This translates to a relative capacity increase of 9.1 % at high intensities and up to 10.3 % at low intensities, with an overall average of 9.6 % (Table 4).

**Table 4: Differences in capacity with and without traffic harmonisation, expressed as the increase of intensity corresponding to specific breakdown probability levels.**

| Breakdown probability | 0.1 % | 0.5 % | 1 % | 2 % | 5 % | 10 % |
|---|---|---|---|---|---|---|
| Corresp. TF intensity [PCE/3 min] w/o VSL | 52.6 | 66.8 | 74.1 | 82.1 | 94.3 | 104.9 |
| Corresp. TF intensity [PCE/3min] with VSL | 58.1 | 73.4 | 81.2 | 89.9 | 103.0 | 114.4 |
| Absolute capacity increase [PCE/3min] | 5.4 | 6.6 | 7.2 | 7.8 | 8.7 | 9.5 |
| Relative capacity increase [%] | 10.3 % | 9.9 % | 9.7 % | 9.5 % | 9.2 % | 9.1 % |

The estimated capacity distributions, modelled by Weibull distributions with parameters estimated using the new MLE formula, are $C \sim W(146.42;\ 6.75)$ without harmonisation and $C \sim W(158.78;\ 6.86)$ with harmonisation. This corresponds to an increase in the median capacity of about 8.1 % (138.7 vs. 150.5 PCE/3min). However, due to the slightly different shape parameter and the fact that such high intensities are never reached in practice, since breakdown typically occurs much earlier, the effective, practical increase in capacity is closer to 9.6 %, as noted above. In fact, the capacity models should not be considered valid beyond the intensity range for which they were developed. While these capacity distributions are specific to the studied work zone, the positive impact of speed harmonisation via VSL can be reasonably expected for any 2-to-1 lane reduction and likely extends to other types of bottlenecks as well.

## 5 Discussion and conclusions

### 5.1 Limitations of existing methods

As demonstrated in Section 4.1, both the original maximum likelihood estimation (MLE) formula from Brilon et al. (2005) and the Kaplan-Meier estimate (PLM), provide qualitatively similar results, but are unsuitable for estimating stochastic capacity. Both methods systematically underestimate breakdown probability at low intensities and overestimate it at high intensities. This is because traffic flow (TF) intensity, unlike time, does not progress linearly, and capacity cannot be observed directly.

Although MLE is in principle a suitable method for this application, the issue lies in the incorrect likelihood function used in past studies. The “old” formula contains a derivation error, possibly introduced to match the results of PLM, which is fundamentally unsuitable for this application. A correct MLE formula was derived and applied to two empirical datasets. The results, compared with those obtained using the PLM and old MLE, clearly demonstrate the advantage of the revised MLE when validated against observed data.

See Sections 3.2 and 3.3 for more details on the discussed methods and their issues.

### 5.2 Case study: impact of speed harmonisation on capacity

The case study using two datasets from the same 2-to-1 lane drop work zone – with and without mobile traffic flow harmonisation system using variable speed limits (VSL) – illustrated the practical applications of stochastic capacity estimation and the impacts of VLS on capacity (Section 4.2). The speed harmonisation improved capacity by almost 10 % or alternatively reduced breakdown probability at the same TF intensity by 40-50 %.

While specific impacts are largely site-dependent, varying positive effects of VSL can be expected even at all types of bottlenecks and even standard dual carriageway sections. This aligns with prior studies (Geistefeldt, 2011; Strömgren & Lind, 2016; Vadde et al., 2011) which show that VSL reduce speed variance and thus reduce traffic disturbances and the risk of breakdowns and accidents.

However, increasing capacity has diminishing returns, as it permits higher TF intensity (or long-term demand), which may once again raise breakdown probability. In this study, the number of free-flow intervals with intensity above 45 PCE/3 min increased by 28 % with the harmonisation across nearly identical observation periods. Meanwhile, the number of recorded breakdowns dropped by 33 % (from 52 to 39), highlighting the benefits of the VSL. However, this is smaller reduction than the almost 50-% decrease in breakdown probability would suggest.

### 5.3 Methodological considerations

TF measurements and capacity definitions in this study were based on overlapping three-minute intervals, evaluated each minute. Consequently, breakdown probabilities must either be assessed on a minute-by-minute basis using the preceding three-minute intensity or transformed using eq. (21) to ensure valid predictions.

Empirical findings also show that relatively minor differences in the Weibull distribution’s scale and shape parameters (with and without harmonisation) can lead to substantial changes in estimated breakdown probabilities. This contrasts with Brilon et al. (2005), which reported shape parameters in the range of 9-15 (using the incorrect MLE formula) and suggested fixing the shape parameter at a mean value of 13. While the exact values are largely affected by the choice of aggregation interval, this approach seems questionable and could lead to considerable over- or underestimation of breakdown probability across different parts of the intensity range. On the other hand, the difference in the shape parameter in this study is indeed minuscule (6.75 vs. 6.86), so fixing it may not cause significant issues when using the corrected MLE formula. This could then simplify capacity predictions for planned work zones or freeway segments, allowing to use simpler regression models to modify only the scale parameter based on their known characteristics, similar to regression models for deterministic capacity. The effect of fixing the shape parameters, at least within certain basic type of road layout, remains to be rigorously evaluated. However, the differences in shape parameter cannot be expected to remain insignificant for different work zone or freeway layouts.

The choice of Weibull distribution in this study was based on earlier works (Brilon & Zurlinden, 2003; Chao et al., 2013), although their findings were based on the old MLE formula and may not hold true under the corrected version. Other studies, such as (Elefteriadou et al., 2011), suggest that a log-normal distribution may also offer a good fit, though without a direct comparison to the Weibull. In (Weng & Yan, 2016), log-normal distribution was used to estimate deterministic capacity of a planned work zone based on known layout and traffic composition. Further research is needed to determine the most appropriate distribution for stochastic capacity modelling in various use cases.

There are some limitations of this study, primarily related to data reliability and interpretative uncertainty. Assigning a specific TF intensity as the direct cause of a breakdown is inherently challenging, especially when data is missing at critical moments or when congestion spillback from downstream sections (e.g., the work zone) cannot be ruled out.

Whether such spillbacks should be attributed to the upstream bottleneck remains open to interpretation. It might be more appropriate to treat and model these cases as phantom congestions in the single-lane segment, since whether (and when) spillback reaches the upstream detector depends on the location and severity of the original disturbance. However, this would require reliable detection of spillbacks from the work zone and separate phantom congestion model, potentially also based on stochastic capacity, and would result in even more complex congestion prediction model.

These uncertainties, along with arbitrary threshold definitions, however, reflect limitations of the data rather than flaws in the estimation method itself. As such, they do not invalidate the conclusion that PLM is unsuitable, and that MLE is appropriate only when based on a correctly defined likelihood function.

### 5.4 Practical applications

The stochastic capacity distributions can be applied in various use cases, including traffic simulation with stochastic breakdowns, a priori travel time predictions using Monte Carlo simulations, or in intelligent traffic control systems. However, predicting the exact moment of breakdown remains inherently challenging due to the chaotic nature of TF and the relatively low probability of breakdown at any given moment.

The distributions may also serve as input for other capacity estimation methods that aim to identify optimal TF intensity level, such as the sustained flow index (SFI) proposed by Shojaat et al. (2017) or the traffic efficiency introduced in (Brilon & Zurlinden, 2003). Intelligent traffic control systems can then aim to sustain this optimal flow to minimize the risk of breakdown and capacity drop while maintaining high throughput.

Several papers have employed SFI to estimate optimal volume for a freeway corridor with multiple bottlenecks (Shojaat et al., 2020) or to define design capacities (Shojaat et al., 2018) or adapted the method to work with TF density instead of intensity (Geistefeldt & Shojaat, 2019). However, they all relied on the incorrect MLE formula, making the numerical results (not the methods) unreliable, along with the uncertainty discussed in the following paragraph.

SFI outputs different optimal TF intensity depending on the aggregation (or rather failure testing) interval. For short intervals, the recommended values are unsustainable. With growing interval length, the recommended intensity decreases (asymptotically towards zero for infinite interval duration). This not necessarily an error – it can be interpreted as optimizing for different time frame, i.e., very high intensity is optimal if optimizing for a short-term window (e.g., a queue of vehicles behind a moving bottleneck) while much lower intensity is optimal for the peak hour, and even lower intensity is optimal for whole-day operation with (mostly) constant demand. However, the choice of optimization is defined by the aggregation (or failure testing, if different) interval, unless the survival probabilities are transformed via eq. (22) and the method does not give any answer to what optimization interval should be used in given moment.

New models of optimal traffic flow intensity to address this issue should be developed, incorporating the expected traffic demand during a longer period (e.g., the rest of the day), and the negative impact of capacity drop. It might be beneficial to put more weight on sustaining free-flow conditions to prevent the capacity drop and consequent queue build-up. Metering ITS systems should then aim to maintain that TF intensity to reach optimal performance – allowing high throughput while keeping a reasonably low risk of breakdown and consequent congestion and queue build up. However, this requires some suitable buffer area to store the excess traffic.

Stochastic capacity can also be used to model queue discharge flow. In combination with stochastic traffic demand model such as exponential distribution of arrivals, queue onset, growth, and dissipation can be modelled fully stochastically.

## 5.5 Additional future research

Further research should focus on expanding the capacity model with additional variables affecting the breakdown probability. Another topic is the reliability of the capacity estimates with respect to the amount of available breakdown data, which is particularly relevant for estimating capacity of work zones, which are only temporary and may even change layout. Gathering more data from additional work zones and road segments to search for the best-fitting distribution is both also relevant. Different capacity definitions and data processing methods can also lead to design of more customised measurements that will better fit the needs of the stochastic capacity estimation for specific locations. Suitable thresholds and definitions for different layouts and applications should also be further studied and discussed.

## Acknowledgements

The author would like to thank those who provided feedback on all versions of the manuscript, including the anonymous reviewers.

This article was produced with financial support of the Czech Ministry of Transport within the programme of long-term conceptual development of research institutions.

While the Transportation Research Centre was involved in the original ViaZONE project, it does not benefit from the implementation of ZIPMANAGER and/or its new iterations.

## References

Ambrožič, M., & Gorjan, L. (2011). Reliability of a Weibull analysis using the maximum-likelihood method. *Journal of Materials Science*, *46*(6), 1862–1869. https://doi.org/10.1007/s10853-010-5014-2

Arnesen, P., & Hjelkrem, O. A. (2018). An Estimator for Traffic Breakdown Probability Based on Classification of Transitional Breakdown Events. *Transportation Science*, *52*(3). https://doi.org/10.1287/trsc.2017.0776

Banks, J. H. (1990). Flow Processes at a Freeway Bottleneck. *Transportation Research Record*, *1287*, 20–28.

Brilon, W., Geistefeldt, J., & Regler, M. (2005). Reliability of Freeway Traffic Flow: A stochastic Concept of Capacity. *Proceedings of the 16th International Symposium on Transportation and Traffic Theory*, 125–144. https://www.ruhr-uni-bochum.de/verkehrswesen/download/literatur/ISTTT16_Brilon_Geistefeldt_Regler_final_citation.pdf

Brilon, W., Geistefeldt, J., & Zurlinden, H. (2007). Implementing the concept of reliability for highway capacity analysis. *Transportation Research Record*, *2027*, 1–8. https://doi.org/10.3141/2027-01

Brilon, W., & Zurlinden, H. (2003). Überlastungswahrscheinlichkeiten und Verkehrsleistung als Bemessungskriterium für Straßenverkehrsanlagen. *Schriftenreihe Forschung Straßenbau Und Straßenverkehrstechnik*, *870*, 108.

Cassidy, M. J., & Bertini, R. L. (1999). Some traffic features at freeway bottlenecks. *Transportation Research Part B: Methodological*, *33B*(1), 25–42. https://doi.org/10.1016/S0191-2615(98)00023-X

Chao, Y., Zhengzheng, C., & Chenrui, H. (2013). Influence Factor Analysis of Stochastic Capacity of Shanghai Expressway Segment. *Procedia - Social and Behavioral Sciences*, *96*, 2497–2505. https://doi.org/10.1016/j.sbspro.2013.08.279

Chow, A. H. F., Lu, X.-Y., & Qiu, T. Z. (2009). An empirical analysis of traffic breakdown. *2nd International Symposium on Freeway and Tollway Operations*, *June*. https://doi.org/10.13140/2.1.2850.6880

Chung, K., Rudjanakanoknad, J., & Cassidy, M. J. (2007). Relation between traffic density and capacity drop at three freeway bottlenecks. *Transportation Research Part B: Methodological*, *41*(1), 82–95. https://doi.org/10.1016/j.trb.2006.02.011

Čičić, M., & Johansson, K. H. (2022). Front-tracking transition system model for traffic state reconstruction, model learning, and control with application to stop-and-go wave dissipation. *Transportation Research Part B: Methodological*, *166*, 212–236. https://doi.org/10.1016/j.trb.2022.10.008

Cicic, M., Mikolasek, I., & Johansson, K. H. (2020). Front tracking transition system model with controlled moving bottlenecks and probabilistic traffic breakdowns. *IFAC-PapersOnLine*, *53*(2), 14990–14996. https://doi.org/https://doi.org/10.1016/j.ifacol.2020.12.1997

Daamen, W., Buisson, C., & Hoogendoorn, S. P. (2014). Traffic simulation and data: Validation methods and applications. In *Traffic Simulation and Data: Validation Methods and Applications*. CRC Press. https://doi.org/10.1201/b17440

Elefteriadou, L., Kondyli, A., Washburn, S., Brilon, W., Lohoff, J., Jacobson, L., Hall, F., & Persaud, B. (2011). Proactive ramp management under the threat of freeway-flow breakdown. *Procedia - Social and Behavioral Sciences*, *16*, 4–14. https://doi.org/10.1016/j.sbspro.2011.04.424

Elefteriadou, L., & Lertworawanich, P. (2003). Defining, Measuring and Estimating Freeway Capacity. *82nd Annual Transportation Research Board Meeting, 12-16 January, Washington D.C.*

Elefteriadou, L., Roess, R. P., & McShane, W. R. (1995). Probabilistic nature of breakdown at freeway merge junctions. *Transportation Research Record*, *1484*, 80–89.

Elefteriadou, L., Torbic, D., & Webster, N. (1996). Development of passenger car equivalents for freeways, two-lane highways, and arterials. *Transportation Research Record*, *1572*, 51–58.

Gazis, D. C., & Foote, R. S. (1969). Surveillance and Control of Tunnel Traffic by an On-Line Digital Computer. *Transportation Science*, *3*(3), 255–275. https://doi.org/10.1287/trsc.3.3.255

Geistefeldt, J. (2011). Capacity effects of variable speed limits on German freeways. *Procedia - Social and Behavioral Sciences*, *16*, 48–56. https://doi.org/10.1016/j.sbspro.2011.04.428

Geistefeldt, J., & Brilon, W. (2009). Transportation and Traffic Theory 2009: Golden Jubilee. In *Transportation and Traffic Theory 2009: Golden Jubilee*. Springer. https://doi.org/10.1007/978-1-4419-0820-9

Geistefeldt, J., & Shojaat, S. (2019). Comparison of Stochastic Estimates of Capacity and Critical Density for U.S. and German Freeways. *Transportation Research Record*, *2673*(8), 388–396. https://doi.org/10.1177/0361198119843471

Greenshields, B. D., Bibbins, J. R., Channing, W. S., & Miller, H. H. (1935). A study of traffic capacity. *Highway Research Board Proceedings*.

Hall, F. L., & Agyemang-Duah, K. (1991). Freeway Capacity Drop and the Definition of Capacity. *Transportation Research Record*, *1320*, 91–98.

Hyde, T., & Wright, C. C. (1986). Extreme value methods for estimating road traffic capacity. *Transportation Research Part B*, *20*(2), 125–138. https://doi.org/10.1016/0191-2615(86)90003-2

Hymel, K. (2019). If you build it, they will drive: Measuring induced demand for vehicle travel in urban areas. *Transport Policy*, *76*, 57–66. https://doi.org/10.1016/j.tranpol.2018.12.006

Kaplan, E. L., & Meier, P. (1958). Nonparametric Estimation from Incomplete Observations. *Journal of the American Statistical Association*, *53*(282), 457–481. https://doi.org/10.1080/01621459.1958.10501452

Kianfar, J., & Abdoli, S. (2021). Deterministic and Stochastic Capacity in Work Zones: Findings from a Long-Term Work Zone. *Journal of Transportation Engineering, Part A: Systems*, *147*(1). https://doi.org/10.1061/JTEPBS.0000470

Knoop, V., & Hoogendoorn, S. (2022). Free Flow Capacity and Queue Discharge Rate: Long-Term Changes. *Transportation Research Record: Journal of the Transportation Research Board*, *2676*(7), 483–494. https://doi.org/10.1177/03611981221078845

Litman, T. (2017). *Generated traffic and induced travel*. Victoria Transport Policy Institute. https://www.vtpi.org/gentraf.pdf

Lorenz, M. R., & Elefteriadou, L. (2001). Defining freeway capacity as function of breakdown probability. *Transportation Research Record*, *1776*, 43–51. https://doi.org/10.3141/1776-06

Mikolasek, I. (2020). *New stochastic highway capacity estimation method and why product limit method is unsuitable*.

Minderhoud, M. M., Botma, H., & Bovy, P. H. L. (1997). Assesment of Roadway Capacity Estimation Methods. *Transportation Research Record*, *1572*(1), 59–67. https://doi.org/https://doi.org/10.3141/1572-08

Myung, I. J. (2003). Tutorial on maximum likelihood estimation. *Journal of Mathematical Psychology*, *47*(1), 90–100. https://doi.org/10.1016/S0022-2496(02)00028-7

Pollak, R. D., & Palazotto, A. N. (2009). A comparison of maximum likelihood models for fatigue strength characterization in materials exhibiting a fatigue limit. *Probabilistic Engineering Mechanics*, *24*(2), 236–241. https://doi.org/10.1016/j.probengmech.2008.06.006

Ščerba, M., Apeltauer, T., & Apeltauer, J. (2015). Portable telematic system as an effective traffic flow management in workzones. *Transport and Telecommunication*, *16*(2), 99–106. https://doi.org/10.1515/ttj-2015-0009

Shao, C. Q. (2011). Implementing estimation of capacity for freeway sections. *Journal of Applied Mathematics*. https://doi.org/10.1155/2011/941481

Shojaat, S., Geistefeldt, J., Parr, S. A., Escobar, L., & Wolshon, B. (2017). Applying the Sustained Flow Index to Estimate Freeway Capacity. *TRB 96th Annual Meeting Compendium of Papers*, *1*(225), 1–17. https://trid.trb.org/view/1439720

Shojaat, S., Geistefeldt, J., Parr, S. A., Escobar, L., & Wolshon, B. (2018). Defining freeway design capacity based on stochastic observations. *Transportation Research Record*, *2672*(15), 131–141. https://doi.org/10.1177/0361198118784401

Shojaat, S., Geistefeldt, J., & Wolshon, B. (2020). Optimum Volume of Freeway Corridors. *Transportation Research Record*, *2674*(3), 27–36. https://doi.org/10.1177/0361198120908249

Srivastava, A., & Geroliminis, N. (2013). Empirical observations of capacity drop in freeway merges with ramp control and integration in a first-order model. *Transportation Research Part C: Emerging Technologies*, *30*, 161–177. https://doi.org/10.1016/j.trc.2013.02.006

Strömgren, P., & Lind, G. (2016). Harmonization with Variable Speed Limits on Motorways. *Transportation Research Procedia*, *15*, 664–675. https://doi.org/10.1016/j.trpro.2016.06.056

Transportation Research Board. (2016). *Highway Capacity Manual - A Guide for Multimodal Mobility Analysis* (6th ed.). Transportation Research Board. http://www.trb.org/Main/Blurbs/175169.aspx

Vadde, R., Sun, D., Sai, J. O., Faruqi, M. A., & Leelani, P. T. (2011). A simulation study of using active traffic management strategies on congested freeways. *Journal of Modern Transportation*, *19*(3), 178–184. https://doi.org/10.3969/j.issn.2095-087X.2012.03.008

Van Toorenburg, J. A. C. (1986). *Praktijkwaarden voor de capaciteit: proefondervindelijke vaststelling van de capaciteit op enkele volledig belaste wegvakken op 2 x 2-en 2 x 3-strooks autosnelwegen en op een enkelbaansweg*. Ministerie van Verkeer en Waterstaat, Rijkswaterstaat, Dienst Verkeerskunde.

Waleczek, H., Geistefeldt, J., Cindric-Middendorf, D., & Riegelhuth, G. (2016). Traffic Flow at a Freeway Work Zone with Reversible Median Lane. *Transportation Research Procedia*, *15*, 257–266. https://doi.org/10.1016/j.trpro.2016.06.022

Weng, J., & Yan, X. (2016). Probability distribution-based model for work zone capacity prediction. *Journal of Advanced Transportation*, *50*(2), 165–179. https://doi.org/10.1002/atr.1310

Zhang, L., & Levinson, D. (2004). Some properties of flows at freeway bottlenecks. *Transportation Research Record*, *1883*(1883), 122–131. https://doi.org/10.3141/1883-14

**Appendix A – Detailed data processing procedure**

1. Filter the raw event-based data to remove invalid records (as indicated by the radar), duplicate entries, and obvious errors.
2. Aggregate the data into one-minute intervals. Calculate speed as a harmonic mean to better reflect the spatial average speed (space mean speed) (Daamen et al., 2014). Vehicles longer than 9 m are considered equivalent to two passenger cars; therefore, TF intensity is expressed in passenger car equivalents (PCE). For details on PCE estimation, see Elefteriadou et al. (1996) or HCM (Transportation Research Board, 2016).
3. Aggregate the one-minute intervals into overlapping three- and five-minute intervals. Sum the vehicle counts and calculate the speed as the arithmetic mean of the harmonic means to approximate the space-time mean speed.
4. Identify breakdown as three-minute interval with an average speed below 40 km/h.
5. Find the first one-minute interval with speed below 40 km/h. The TF intensity during the three-minute interval immediately preceding that minute is considered as the breakdown flow unless the data suggest that the queue had already begun forming in the previous minute.
6. Record the breakdown flow intensity (only one value per event) as uncensored observation, and mark all (see step 8 for exceptions) the overlapping free-flow intervals before it as censored data (see Section 3.2).
7. Identify the end of the congestion as an average speed exceeding 70 km/h over a five-minute interval.
8. Continue searching for additional breakdowns. Use all subsequent free-flow data (before the next breakdown) as censored observations. Data during congestion and intervals with brief, inconclusive speed drops below 50 km/h or with intensity below 45 PCE/3 min are discarded.
9. Repeat steps 5-8 until the end of the dataset. After the final breakdown, record the remaining free-flow data as censored observations.